\documentclass{article}

\usepackage{lineno,hyperref}
\usepackage{lipsum}
\usepackage{mathtools, cuted}
\usepackage{authblk}
\modulolinenumbers[1]
\usepackage[left=1.7cm,right=1.7cm,top=4cm,bottom=4cm]{geometry}
\usepackage{multicol}
\usepackage{xcolor}
\usepackage{array}
\usepackage{cite}

\title{Exact expression for the propagating front velocity in nonlinear discrete systems under nonreciprocal coupling}

\author[1]{David Pinto-Ramos}
\date{}

\affil[1]{\it Center for Advanced Systems Understanding (CASUS); Helmholtz-Zentrum Dresden-Rossendorf (HZDR), Görlitz, Germany}
\begin{document}
    \maketitle
    \begin{abstract}
   Nonlinear waves are a robust phenomenon observed in complex systems ranging from mechanics to ecology. Fronts are fundamental due to their robustness against perturbations and capacity to propagate one state over another. Controlling and understanding these waves is then fundamental to make use of their properties. Their velocity is one of the most important properties, which can be theoretically computed only in limited conditions of the dynamical system, and it becomes elusive in the presence of spatial discreteness and nonreciprocal coupling. This work reveals that fronts in discrete systems can be treated as rigid objects when analyzing their whole trajectory instead of the instantaneous one. Then, a relationship between the front velocity and its found shape is given. The formula provides insight into fronts' long-observed properties and agrees with the approximative and parameterized methods described in the literature. Numerical simulations show perfect agreement with the theory.
    \end{abstract}

\subsection*{Introduction} 

\begin{figure}[ht]
	\centering
	\includegraphics[width=0.85\columnwidth]{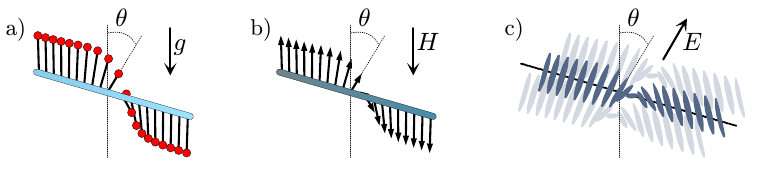}
	\caption{Schematic cartoon of different systems exhibiting front dynamics. a) chain of pendulums subjected to gravity $g$, the angle $\theta$ measures the pendulum angle with respect to the vertical axis. The upside-down position is an unstable equilibrium that connects with the stable upright position through a front. b) an array of classical magnets subjected to a magnetic field forcing $H$. The coupling that aligns each one with its neighbors creates two possible equilibrium positions, measured by $\theta$. The magnetic field makes one of these positions prefered over the other. A front connects these two equilibria, forming a so-called \textit{domain wall}. c) a nematic liquid crystal exhibiting orientational order subjected to an electric field $E$. If the liquid crystal is subjected to planar strong anchoring (with walls perpendicular to the electric field showed) and has a positive dielectric anisotropy (the molecules pictured want to align parallel or antiparallel electric field direction), then two possible values of the angle $\theta$ describing the orientation of the molecules are stable. A front connects these equilibria.}
	\label{F0}
\end{figure}

Interfaces separating two states of a complex system are ubiquitous in nature, being observed in mechanical systems \cite{alfaro2017pi,pintoramos2021nrcisa,veenstra2024non}, fluids \cite{pomeau1986front,ahlers1983vortex,fineberg1987vortex}, magnetic devices \cite{slonczewski1973theory,metaxas2007creep,braun2013frenkel}, granular media \cite{jara2020noise,crespin2021particle}, liquid crystals experiments \cite{alvarez2019front,alfaro2020front,alfaro2018front,residori2005patterns}, and even ecological models \cite{zelnik2017desertification,fernandez2019front}, to mention a few. They result from the system displaying multiple equilibria for an order parameter and being spatially coupled to neighbors. Different equilibria can be established in the extended system at different spatial locations, and an \textit{interface} emerges to reconcile the otherwise abrupt change of the order parameter smoothly, mediated by the spatial coupling mechanism. Some schematic examples of this phenomenon are illustrated in Fig. \ref{F0}, where extended systems of the mechanical (a), magnetic (b), and liquid crystal (c) type can be described by a single order parameter ($\theta$) as a function of space and time, which exhibits two possible equilibrium configurations and an interface separates them. 

The interfaces separating two equilibria are hardly static, and unless in the presence of specific conditions, they will show some degree of motion. A propagating interface is often called a \textit{front}. These structures have been thoroughly studied due to their robustness against perturbations, and controlling their propagation has become a subject of its own. Typical applications correspond to the transmission of information and energy, which is encoded in the different equilibrium states of the nonlinear system and is transmitted thanks to the front propagation \cite{allwood2005magnetic,veenstra2024non,kochmann2017exploiting,nadkarni2016unidirectional,raney2016stable}. Thus, control mechanisms have been explored, the most simple one corresponding to a driving force that favors one state over the other, as seen in Fig. \ref{F0} (b), where a magnetic field drives the system. This mechanism controls the propagation speed but not its direction, as the favored state (the one with the least free energy) will propagate isotropically. A mechanism to control both the speed and direction of these fronts has been proposed recently theoretically \cite{pintoramos2021nrcisa}, and later proven experimentally \cite{aguilera2024nonlinear,veenstra2024non}. This novel mechanism consists of controlling the coupling between the elements, making it \textit{nonreciprocal}, and it is achieved by sophisticated methods such as constructing metamaterials \cite{brandenbourger2019non,veenstra2024non} or employing optical feedback \cite{aguilera2024nonlinear}.

Despite these mechanisms being successful in controlling the speed and direction of front propagation, it is unclear how to theoretically predict the front velocity accurately for all parameter values. The reported predicted velocities are based on approximations taking the continuum limit for the order parameter \cite{veenstra2024non,aguilera2024nonlinear}, or averaging techniques \cite{pintoramos2021nrcisa}. However, doing so leaves aside unexpected phenomena of front propagation, such as the \textit{hopping dynamics} (oscillations while traveling) observed for the position of the front \cite{alfaro2019traveling, alfaro2017pi}, or the propagation pinning phenomenon (or propagation failure) \cite{cahn1960theory,kladko2000universal,braun2013frenkel,clerc2011continuous,pinto2023giant}. Despite the efforts made to account for the front velocity in the presence of spatial discreteness effects, it has been an elusive problem due to the difficulties in addressing nonlinear difference equations. Several approaches have been employed to provide a formula for the velocity. Still, they rely either on a parameterized effective description of the discreteness \cite{alfaro2017pi,clerc2011continuous}, or a weakly discrete limit to predict the front spatial form (in most cases required to compute the velocity) via an effective partial differential equation \cite{flach1996perturbation,kladko2000universal}.

In this work, I intend to provide a parameterization-free, exact description of the front velocity, employing solely the discrete extended system equations and numerical analysis. The description of the method is focused on fronts connecting two stable states, but it is easily generalizable to other situations. For this, I first show that the front spatial shape can be obtained at all times from a spatially continuous function. Having identified the front shape, simple mathematical manipulations allow one to provide a formula for the front velocity that captures its oscillatory nature, depends only on the discrete system, and is valid for any parameter value. I employ the front spatial shape computed numerically in the velocity equation, allowing me to obtain the exact velocity of the front. This is verified with numerical simulations of a prototypical model of bistable nonlinear systems under nonreciprocal coupling.

\subsection*{Theoretical model}

\subsubsection*{A generic dynamical system of fronts}

The simplest model that can generate fronts between stable states requires an array of dynamical systems exhibiting bistability and coupling to neighbors. The universal imperfect pitchfork normal form describes the individual bistable nonlinear systems, allowing for a bias towards one of the stable equilibria \cite{strogatz2018nonlinear}. Simple nearest neighbors coupling is employed and will be promoted to be nonreciprocal, with the scope of describing the new mechanism that controls both the speed and direction of fronts. Under these considerations, the generic model for fronts reads
\begin{eqnarray}
	\dot{A}_k= \eta + \epsilon A_k - A_k^3 + (D-\alpha)(A_{k+1}-A_k) -(D+\alpha)(A_k- A_{k-1}). \label{eq1}
\end{eqnarray}

Similar equations have been derived explicitly in a liquid crystal experiment with optical feedback and for nonlinear oscillators coupled with robotic metamaterials \cite{veenstra2024non,aguilera2024nonlinear}. The first three terms represent the bistable dynamics of a single element: $\eta$ is a bias towards one of the equilibria, $\epsilon$ is the linear grow term, and $A^3$ is the minimal saturation. The fourth and fifth terms represent the coupling with the forward and backward elements, respectively. $D$ is the reciprocal coupling constant, and $\alpha$ is the nonreciprocal one. 

\subsubsection*{Definition of the front position}

\begin{figure}[]
	\centering
	\includegraphics{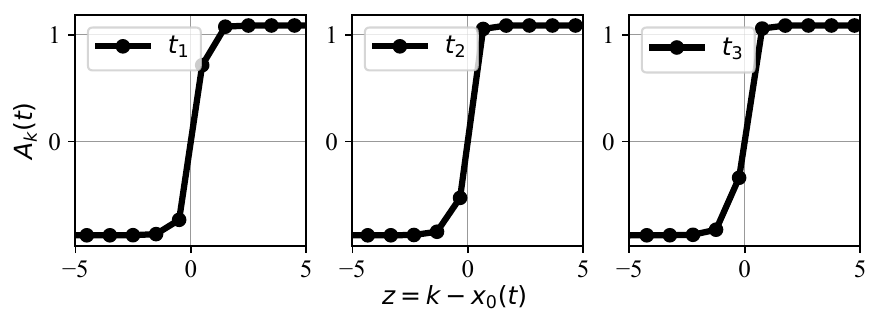}
	\caption{Fronts into the stable state obtained from numerical integration of Eq.~\ref{eq1}. The three panels show the solutions $A_k(t)$ as a function of $z=k-x_0(t)$ at 20 simulation time unit intervals ($t_1<t_2<t_3$). Parameters correspond to $\eta= 0.2$, $\epsilon= 1$, $D=0.1$, $\alpha=-0.05$.}
	\label{F1}
\end{figure}

To analyze the velocity of the front, first one needs to define its position.  A straightforward definition is to set a threshold value $A_T$ such that if $A_{i_0}(t)<A_T<A_{i_0+1}(t)$, then the front position $x_0$ is defined as
\begin{eqnarray}
	x_0 =  i_0+ \frac{A_T-A_{i_0}(t)}{A_{i_0+1}(t)-A_{i_0}(t)}.
\end{eqnarray} 
This is just the continuous value $x_0$ at which the linear interpolated profile $A_i(t)$ reaches $A_T$. The front profile at different times as a function of $z=k-x_0(t)$ is seen in figure \ref{F1}. One notices that, apparently, the front does not translate as a rigid object. One can choose any value of $A_T$ between the values of the equilibria. I selected $A_T=0$ for simplicity.

\subsection*{Results}
\subsubsection*{The continuous profile characterizing the discrete front}  

Snapshots of the system at different times would convince oneself that the front is not a rigid object, unlike what one observes in the continuum limit of Eq. \eqref{eq1} where the translation invariance allows for any continuous translation of the front shape to be a solution as well. Similarly, the system under study has a discrete translation invariance, corresponding to translating the system by one unit cell; thus, the front profile has to undergo some oscillatory motion to achieve this. One may ask oneself the question: Could the solutions at any time $A_k(t)$ be sampled from a simple curve? This would be a significant property, as most methods to compute the front velocity rely on a change of variables such that the velocity becomes explicit. 

\begin{figure}[]
	\centering
	\includegraphics[width=0.8\columnwidth]{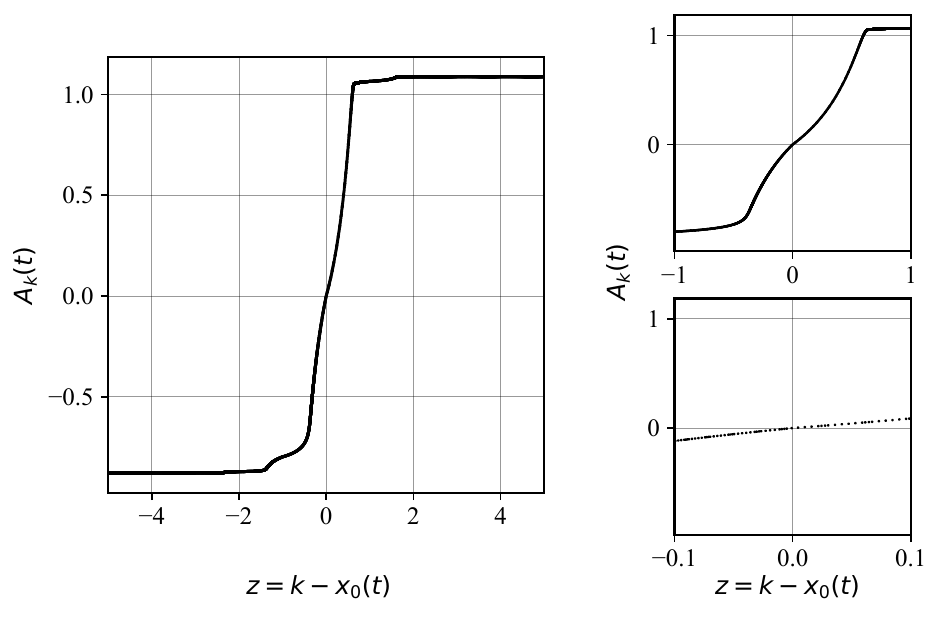}
	\caption{The continuous front profile obtained by collecting the solution points $(z, A_k(t))$. The left panel shows the profile around the position of the front, and the right insets show more detail up to the points forming the continuous function; 500 snapshots were used to construct this figure. Parameters correspond to $\eta= 0.2$, $\epsilon= 1$, $D=0.1$, $\alpha=-0.05$.}
	\label{F2}
\end{figure}

Let me define the variable $z=k-x_0(t)$, corresponding to centering the front in its position and using $z$ as the new indexes for the system elements. At each time instant, one has a collection of $z$ numbers (with separation 1 between them) representing the position of the elements measured from the front position. As time evolves, the collection of $z$ numbers does so. The variable $z$ varies continuously, as it is a function of time. Now, one can collect the points $z(t)$ and the system state $A_k(t)$ at each time step and store them. This corresponds to the system state at all previous times. If enough time has elapsed, one will have collected a high number of points, corresponding exactly to $NT$, with $N$ the number of elements in Eq. \eqref{eq1} and $T$ the time steps. Then, one can plot them in the $(A, z)$ plane. Unexpectedly, the points form a smooth curve in this plane, and they reconstruct a well-defined function describing the front at all times. This is seen in Fig. \ref{F2}. This result means that if the position of the front is known, one could obtain the $z$ coordinates of the system elements and then draw the system state from the curve depicted in Fig. \ref{F2}. Let me call this curve the \textit{front profile} $A(z)$. In what follows, I will ignore the boundary effects, then, analytical calculations will be carried out assuming an infinite array of elements.

\subsubsection*{The front velocity expression}

The fact that a rigid-shaped front exists allows one to perform the formal change of variables to the co-mobile reference frame $z=k-x_0(t)$ \cite{van2003front}

$$\dot{A}_k(t) = -\dot{x}_0 A'(z).$$

Then, the $N$ equations of motion describing the array of nonlinear elements satisfy the equation

\begin{eqnarray*}
	-\dot{x}_0 A'(z) &=& (\eta + \epsilon A(z) -A^3(z)) + (D-\alpha)(A(z+1)-A(z)) - (D+\alpha)(A(z)-A(z-1)), \\
	&=& f\left(A(z)\right) + (D-\alpha)\left(A(z+1)-A(z)\right) - (D+\alpha)\left(A(z)-A(z-1)\right).
\end{eqnarray*}

Finally, one can sum over the elements to obtain a function of time alone. Thus, one multiplies by $A'(z)$ and sum over the $N$ elements as follows

\begin{eqnarray}
	-\dot{x}_0 = \frac{	\sum_k \left[ f\left(A(z)\right) + (D-\alpha)(A(z+1)-A(z)) - (D+\alpha)(A(z)-A(z-1)) \right] A'(z)}{\sum_k \left(A'(z)\right)^2}. \label{vel_formula}
\end{eqnarray}

One limitation is that the formula depends on the rigid front profile $A(z)$. However, this is not unexpected, as it is usually the case even in continuous systems \cite{pismen2006patterns}. To obtain a closed expression, one needs to compute the front profile analytically, which is out of the scope of this work.

\subsubsection*{Fourier series representation}

\begin{figure}[]
	\centering
	\includegraphics[width=\columnwidth]{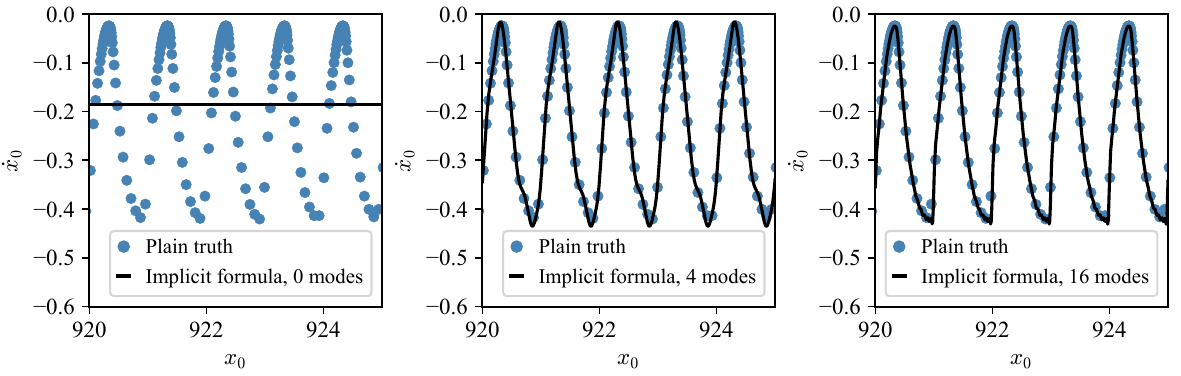}
	\caption{Velocity of the front as a function of the position $\dot{x}_0(x_0)$. The graph shows the plain truth, the implicit formula \eqref{vel_boni} using 0, 4, and 16 terms in the Fourier series representation. Parameters correspond to $\eta= 0.2$, $\epsilon= 1$, $D=0.1$, $\alpha=-0.05$.}
	\label{Front_Velocity_Comp}
\end{figure}

One can explore the expression \eqref{vel_formula} further. One can note that the front profile $A(z=k-x_0(t))$ is invariant for the transformation $k\rightarrow k+1$ and $x_0(t)\rightarrow x_0(t+\tau) = x_0(t) +1$ for some $\tau$. $\tau$ is the time the front position takes to advance one element. Now, one needs to consider expressions of the type $\sum_k^N A^m(z=k-x_0(t))$; for a fixed time $t$, one is choosing $N$ points from the continuous profile $A(z)$ and summing them. These $N$ points will repeat (except by the two at the ends of the chain) again after the time $\tau$, or equivalently, when a shift in one cell occurs. Then, the expression \eqref{vel_formula} can be either considered a periodic function of $t$ (of period $\tau$) or a periodic function of $x_0$ (of period one). This allows writing $\dot{x_0}$ in a Fourier series representation; moreover, the numerator and denominator in Eq.~\eqref{vel_formula} can be expressed in Fourier series separately. The denominator reads
\begin{eqnarray*}
	\left(\sum_k A'(z)^2\right)(x_0) = C_0\left[ A'(z)^2\right] +\sum_{n=1}^\infty \left(C_n \left[ A'(z)^2\right] \cos(2 \pi n x_0)+ S_n\left[ A'(z)^2\right]\sin(2\pi n x_0) \right),
\end{eqnarray*}
which I compactly write
\begin{eqnarray*}
	\left(\sum_k A'(z)^2\right)(x_0) = M_0 + \sum_n M_n.
\end{eqnarray*}
The Fourier coefficients have interesting formulas reading
\begin{eqnarray*}
	C_n \left[ A'(z)^2\right] &=& \int_0^1 \left(\sum_{k=1}^{N} A'(z)^2\right) \cos(2\pi k x_0) dx_0 \nonumber \\
	&=& \sum_{k=1}^{N}  \int_0^1  A'(z)^2 \cos(2\pi n x_0) dx_0 \nonumber \\
	&=& \sum_{k=1}^{N}  \int_{i-1}^{i}  A'(z)^2 \cos( 2\pi n (i-z) ) dz \nonumber \\
	&=& \int_0^N  A'(z)^2 \cos(2\pi n z) dz,
\end{eqnarray*}
and
\begin{eqnarray*}
	S_n \left[ A'(z)^2\right]  &=& \int_0^1 \left(\sum_{k=1}^{N}  A'(z)^2\right) \sin(2\pi n x_0) dx_0 \nonumber \\
	&=& -  \int_0^N  A'(z)^2 \sin(2\pi n z) dz.
\end{eqnarray*}
They could be reduced to integrals of the continuous profile characterizing the front dynamic. Similar expressions follow for the numerator terms. The local term $f(A)=\eta+\epsilon A- A^3$ contributes
\begin{eqnarray*}
	\left(\sum_k f(A) A'(z) \right)(x_0) = C_0\left[ f(A)A'(z)\right] +\sum_{n=1}^\infty \left(C_n \left[f(A)A'(z)\right] \cos(2 \pi n x_0)+\right.
	\left. S_n\left[ f(A)A'(z)\right]\sin(2\pi n x_0) \right). 
\end{eqnarray*}
Note that $ C_0\left[ f(A)A'(z)\right] = -( V(z\rightarrow \infty) - V(z\rightarrow-\infty) ) = -\Delta V$, with $V(A)= -\int^{A}f(x)dx$. The front speed has a contribution proportional to the free energy density, similar to fronts in the continuum limit; however, due to the grid, it has periodic contributions. Then, I compactly write
\begin{eqnarray*}
	\left(\sum_k f(A) A'(z) \right)(x_0) = -\Delta V + \sum_n F_n(x_0).
\end{eqnarray*}
One can compute the rest of the terms as follows, using the fact that $\int_0^1 \sum_k \rightarrow \int_0^N$ (using the periodicity allows to write the coefficients as an integral over the whole array of elements)
\begin{eqnarray*}
	D\int\left[A(z+1)+A(z-1)-2A(z)\right]A'(z) dz&=&2D\int \left[(\cosh \partial_z -1)A(z)\right]A'(z) dz \nonumber \\
	&=& 2D\int\left( \left[\frac{1}{2!}\partial_z^2 + \frac{1}{4!}\partial_z^4+ ...\right]A(z)\right) A'(z) dz \nonumber \\
	&=&2D\left[\frac{1}{2!} \left.(A'(z))^2\right|_0^N - \frac{1}{4! 2}\left.(A''(z))^2\right|_0^N+ ... \right] \nonumber \\
	&=& 0.
\end{eqnarray*}
To obtain the last line, one uses the observed fact that the profile $A'(z)$ and its derivatives vanish at the ends of the front. Similarly
\begin{eqnarray*}
	\alpha \int \left[A(z+1)-A(z-1)\right]A'(z)dz &=& 2\alpha \int \left(\left[\sinh \partial_z \right] A(z)\right) A'(z) dz \nonumber \\
	&=& 2\alpha \int \left[ A'(z) + \frac{1}{3!} A'''(z) +...\right]\partial_z u_F(z) dz \nonumber \\
	&=&2\alpha \int \left[(A'(z))^2 -\frac{1}{3!}(A''(z))^2+
	\frac{1}{5!}(A'''(z))^2- ...\right] dz \nonumber \\
	&\equiv& 2\alpha (M_0 + M_\alpha),
\end{eqnarray*} 
where $M_0 = \int (A'(z))^2 dz$ repeats in the denominator of equation \eqref{vel_formula}. Then, one can write for equation \eqref{vel_formula} the following
\begin{eqnarray}
	-\dot{x}_0 = \frac{-\Delta V-2\alpha(M_0+M_\alpha) + \sum_{n=1} F_n(x_0)}{M_0 + \sum_{n=1} M_n(x_0)}. \label{vel_boni}
\end{eqnarray}
One can compare what is obtained using implicit formula \eqref{vel_boni} using a given number of terms in each summation with the plain truth of computing $\dot{x}_0(t)$ numerically in figure \ref{Front_Velocity_Comp}.
What remains an open question is to obtain an approximation or explicit solution to the continuous front profile $A(z)$ characterizing the discrete cells at all times. If that is possible, then the formula \ref{vel_boni} would become explicit, and the velocity would be written as a function of the system parameters only (in this case $\eta$, $\epsilon$, $D$, and $\alpha$).

\subsection*{Discussion and conclusions}

This work provides a general framework to develop further investigations regarding the front velocity in discrete systems under nonreciprocal coupling. It is remarkable that Eq. \eqref{vel_boni} provides simple intuition about the effect of discreteness and nonreciprocity. Indeed, in the continuum limit and for $\alpha=0$, one recovers the well-known relationship for fronts connecting two stable states  $\dot{x}_0=\Delta V/M_0$ \cite{pismen2006patterns}. The effect of $\alpha$ is readily understood, at least its explicit contribution, making the velocity depend linearly on $\alpha$, a fact that recovers the experimental results of references \cite{aguilera2024nonlinear,veenstra2024non}, providing an exact relationship. Lastly, the effect of discreteness appears explicitly as oscillatory terms in the velocity expression ($F_n$ and $M_n$), which is in agreement with the effective continuous theories \cite{clerc2011continuous}.

The result I highlight the most is finding a continuous, rigid front functional shape from which the solutions are sampled, the function $A(z)$. The existence of this function is what allows this theoretical development. One drawback is that the velocity formula depends on this function, which I could only provide numerically. To obtain this function, a parameterized nonlinear difference equation is to be solved (with free parameter the front velocity $\dot{x}_0$); then, one could provide a closed equation for this velocity. Future studies aim in this direction. 

A generic equation for the nonlinear, nonreciprocally coupled system was employed. Nevertheless, the method is directly applicable to more general models. One interesting issue for future studies is to analyze the effect of nonlinear nonreciprocal interactions, such as the ones that appear in the description of liquid crystals subjected to optical feedback \cite{aguilera2024nonlinear}, where a nonlinear response for the velocity with respect to the nonreciprocity level is observed. This work sets the basis for such future investigations.

To conclude, this work provides a method to characterize the front velocity under general conditions with a parameter and approximation-free scheme. The method is generalizable to more complex one-dimensional extended dynamical systems directly, where the control of front motion has become a key point to develop novel mechanisms of lossless information and energy transmission  \cite{aguilera2024nonlinear,veenstra2024non}.

\subsection*{Acknowledgements}
I thank Karin Alfaro-Bittner, Marcel Clerc, and René Rojas for their support and fruitful discussions. This work was partially funded by the Center of Advanced Systems Understanding (CASUS), which is financed by Germany’s Federal Ministry of Education and Research (BMBF) and by the Saxon Ministry for Science, Culture and Tourism (SMWK) with tax funds on the basis of the budget approved by the Saxon State Parliament.

\end{document}